\documentclass[review,12pt]{elsarticle}

\usepackage{amsmath,amssymb}
\usepackage{graphicx}
\usepackage{array}
\usepackage{booktabs}
\usepackage[hypertexnames=false,hidelinks]{hyperref}
\usepackage{float}
\usepackage{placeins}
\hypersetup{
    pdfauthor={Dandan Chen, Yan Zhao, Xuepeng Chen},
    pdfsubject={PV power forecasting; NWP uncertainty; deep learning robustness},
    pdfkeywords={PV power forecasting, NWP uncertainty, deep learning robustness, explainable AI},
    pdfcreator={LaTeX},
    pdfproducer={}
}

\journal{Results in Engineering}

\begin{document}

\begin{frontmatter}

\title{Robustness of Deep Learning Models for PV Power Forecasting under NWP Forecast Errors: A Spatiotemporal and Physically Interpretable Analysis}

\author[inst1,inst2]{Dandan Chen\corref{cor1}}
\ead{chendandan@cma.gov.cn}
\cortext[cor1]{Corresponding author.}

\affiliation[inst1]{organization={State Key Laboratory of Severe Weather Meteorological Science and Technology, Chinese Academy of Meteorological Sciences},
    addressline={No. 46 Zhongguancun South Street},
    city={Beijing},
    postcode={100081},
    country={China}}

\affiliation[inst2]{organization={Key Laboratory of High Impact Weather (special), China Meteorological Administration},
    city={Changsha},
    postcode={410073},
    country={China}}

\author[inst1]{Yan Zhao}
\author[inst1]{Xuepeng Chen}

\begin{abstract}
Engineering use of AI forecasting models requires not only high nominal accuracy but also predictable behavior under uncertain inputs. In photovoltaic (PV) forecasting, this requirement is especially challenging because numerical weather prediction (NWP) errors are temporally correlated, state dependent, and physically coupled across variables. Existing evaluations, however, often rely on perfect forecast assumptions or simplistic perturbations that do not reflect these characteristics. This study presents a physically constrained robustness evaluation framework based on simulation, using virtual PV power as a controlled response variable to isolate the propagation of input uncertainty from confounders at the plant level. Six representative machine learning and deep sequence models, including PatchTST, GRU, N-HITS, and LightGBM, are evaluated under dynamic NWP perturbations with heteroscedasticity modulated by clear-sky conditions and Erbs reconstruction that preserves radiation consistency. The results show that sequence models provide stronger noise filtering and temporal resilience than a strong tabular baseline under medium to high disturbance regimes. SHapley Additive exPlanations (SHAP) and Integrated Gradients (IG) further support a feature reallocation tendency at the case level, in which predictive reliance shifts from corrupted future forecasts toward more stable historical observations and deterministic physical priors. A Pareto analysis of accuracy under clean conditions, robustness, and computational latency then translates these findings into engineering implications for robustness assessment and model selection under forecast uncertainty.
\end{abstract}


\end{frontmatter}

\section{Introduction}

Reliable engineering use of AI forecasters depends not only on predictive accuracy in nominal conditions, but also on how model behavior changes when upstream inputs become uncertain. Photovoltaic (PV) power forecasting provides a representative testbed for this broader problem because it couples data-driven sequence modeling with imperfect numerical weather prediction (NWP) inputs \cite{wan2015photovoltaic,chen2025economic}. While traditional physical, statistical, and early machine learning methods laid the groundwork \cite{golestaneh2016very,antonanzas2016review,shamsi2021prediction}, modern deep learning (DL) architectures have become increasingly effective at extracting nonlinear spatiotemporal structure from meteorological data \cite{mellit2021deep,li2025efficient,dimitriadis2025deep,liu2023renewable}. Yet the engineering reliability of these models under structured forecast uncertainty remains much less understood than their accuracy under clean conditions.

Nevertheless, the engineering use of these advanced AI models is inherently constrained by numerical weather prediction (NWP) uncertainty \cite{sweeney2020future,severiano2021evolving,ge2024joint}. At forecast issuance time $t_0$, sequence models typically integrate reliable historical observations with subsequent NWP inputs. In real world environments, these future drivers—particularly irradiance and temperature—are highly stochastic due to atmospheric chaos, model resolution limits, and rapidly evolving cloud processes \cite{antonanzas2016review}. This input degradation can propagate nonlinearly through deep network architectures, severely compromising the reliability of downstream power predictions \cite{agoua2018probabilistic,wang2023inherent}.

The algorithmic robustness of AI models against such operational shifts is frequently simplified in existing literature. Many studies evaluate forecasting architectures under idealized perfect forecast settings, while robustness stress tests often rely on simple, independent and identically distributed noise injection. While useful for basic sensitivity checks, these setups fail to capture the complex structure of real NWP errors, including temporal persistence, heteroscedasticity related to sky state, and spatial disparities \cite{verzijlbergh2015improved,andrade2017improving,brester2023evaluating,clauzel2024west}. Furthermore, physically consistent perturbations are required when evaluating data-driven models; for instance, perturbing solar radiation components independently without physical constraints results in scenarios that fail to represent true meteorological dynamics \cite{erbs1982estimation}.

Beyond macroscopic predictive accuracy, a critical gap exists in understanding the internal decision patterns of these black-box architectures when confronted with unreliable future inputs. The algorithmic vulnerability to NWP errors is unlikely to be spatially uniform \cite{wang2025capacity}, but rather coupled with local physical regimes such as cloud variability \cite{carpentieri2023intraday}. While existing literature has extensively investigated AI model vulnerabilities under adversarial attacks \cite{ruan2023vulnerability,wen2023regional,li2025interpretable}, the internal spatiotemporal dependency shifts of deep networks under naturally occurring, correlated NWP errors remain poorly understood. Improving this interpretability is essential if AI robustness is to be discussed in physically meaningful rather than purely statistical terms \cite{kim2025comprehensive,liu2025robust}.

To address these gaps, this study proposes a physically constrained evaluation framework to systematically investigate the robustness of diverse AI architectures, using virtual PV power generated by \texttt{pvlib} \cite{anderson2023pvlib,holmgren2018pvlib} as the target. Across five representative U.S. climate zones, we construct a dynamic error injection mechanism that mimics the statistical structure of real NWP errors, calibrated via empirical HRRR residual statistics. To maintain physical realism, perturbations are applied selectively: GHI is perturbed using an autoregressive process modulated by clear-sky conditions, DNI/DHI are reconstructed through Erbs decomposition, and remaining variables follow deterministic or climatological priors. On this basis, we evaluate the architectural resilience of six benchmark models, including strong tabular methods (LightGBM) \cite{ke2017lightgbm} and advanced deep sequence networks (N-HITS \cite{challu2023nhits}, GRU \cite{cho2014learning}, and PatchTST \cite{nie2023time}).

In summary, this study makes four primary contributions. First, we propose a dynamic, physically consistent NWP error injection framework for AI stress testing that captures temporal autocorrelation and heteroscedasticity while maintaining radiation consistency. Second, we introduce a physically informed feature construction strategy that avoids the uniform perturbation of deterministic variables. Third, we show that algorithmic robustness degradation exhibits distinct spatial heterogeneity linked to physical drivers such as jump intensity in sky state. Fourth, by integrating Explainable AI (XAI) techniques—specifically SHapley Additive exPlanations (SHAP) \cite{lundberg2017unified} and Integrated Gradients (IG) \cite{sundararajan2017axiomatic}—with a Pareto trade-off analysis, we provide an interpretation of predictive reliance reallocation at the case level and discuss its implications for robust model selection in complex physical environments.

\section{Data and Methodology}

This section describes the study regions, the construction of operationally relevant forecast inputs, and the dynamic NWP error injection framework. High-resolution meteorological data from the NREL \cite{sengupta2018national} are used together with virtual PV power generated by \texttt{pvlib} \cite{holmgren2018pvlib} to provide a controlled yet physically consistent forecasting benchmark. In this sense, the benchmark can be viewed as a controlled environment similar to a digital twin: rather than reproducing every operational detail of a specific PV plant, it uses a transparent physical conversion model to isolate how NWP forecast errors propagate through forecasting algorithms. Real PV plant measurements often conflate meteorological uncertainty with effects at the equipment and operation levels, such as inverter clipping, soiling, curtailment, sensor calibration drift, and maintenance-related anomalies. Constructing a virtual plant with \texttt{pvlib} therefore provides a strictly controlled digital setting in which meteorological input uncertainty can be separated from mechanical and electrical confounders at the plant level.

\subsection{Study Regions, Baseline Data, and Sample Setup}

\begin{figure}[H]
    \centering
    \includegraphics[width=0.95\textwidth]{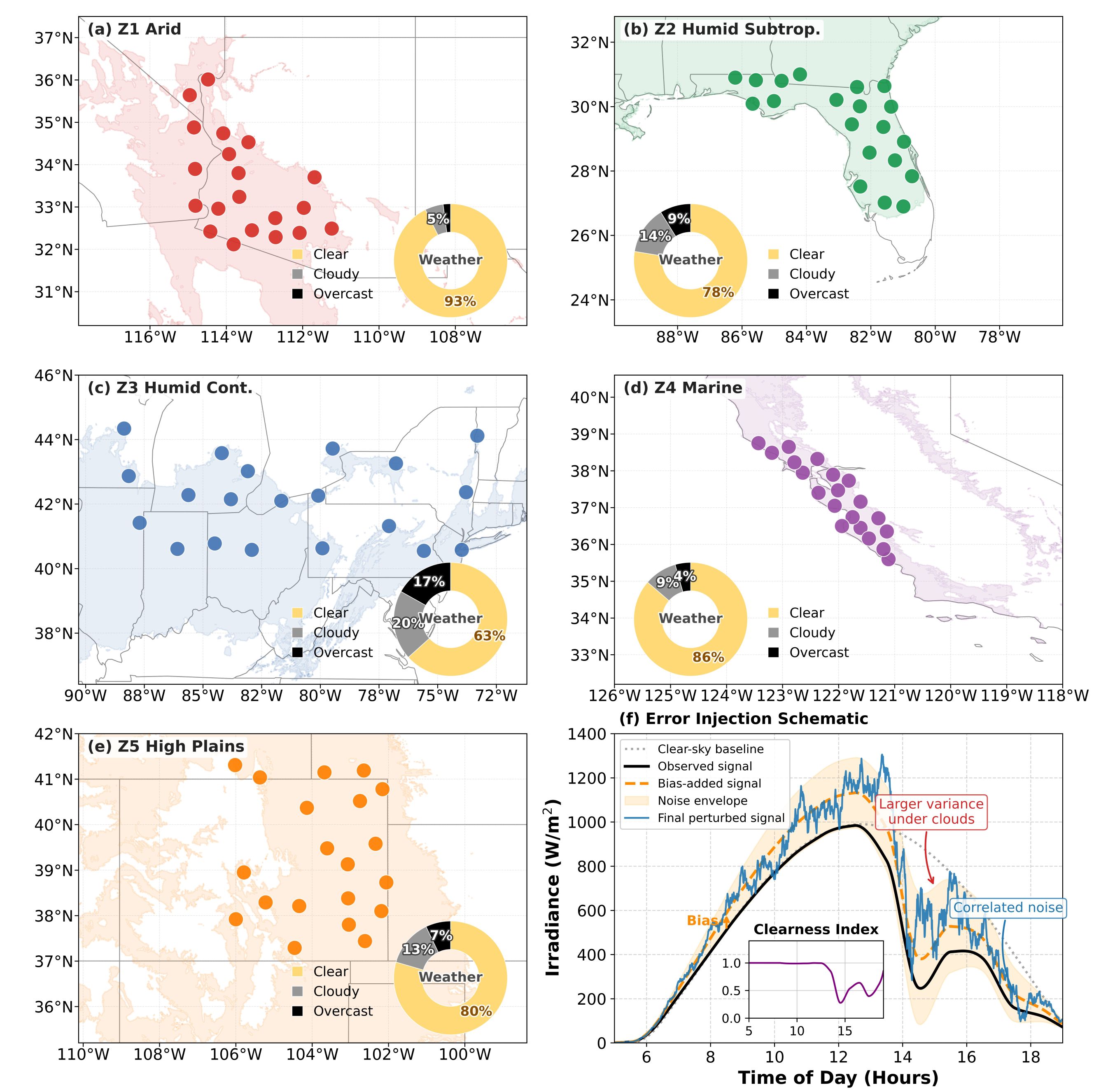}
    \caption{Geographical distribution of the selected climate zones and the proposed error injection methodology. (a)--(e) Spatial mapping of the five diverse climate zones across the contiguous United States: Z1 Arid, Z2 Humid Subtropical, Z3 Humid Continental, Z4 Marine, and Z5 High Plains. The inset charts summarize the climatological proportions of three sky state regimes defined by the observed clearness index: clear ($K_t^{\mathrm{obs}} > 0.65$), cloudy ($0.3 < K_t^{\mathrm{obs}} \le 0.65$), and overcast ($K_t^{\mathrm{obs}} \le 0.3$). (f) Representative illustration of the dynamic error injection framework for an example day (2018-07-08) at a site in the Arid zone (Z1), showing how systematic bias and temporally correlated heteroscedastic noise are superimposed on the clear-sky baseline to generate the final perturbed signal.}
    \label{fig:fig1}
\end{figure}

We consider five representative climate zones across the contiguous United States, namely Z1 Arid, Z2 Humid Subtropical, Z3 Humid Continental, Z4 Marine, and Z5 High Plains, selected according to the Koppen--Geiger classification \cite{kottek2006world,peel2007updated,beck2018present} and NREL site availability. As shown in Fig.~\ref{fig:fig1}, these regions span diverse meteorological regimes, from relatively stable clear-sky environments to highly variable conditions dominated by clouds, thereby providing a natural basis for examining spatial heterogeneity in robustness.

The forecasting target is virtual PV power generated from reference meteorological drivers under a unified \texttt{pvlib} configuration. For each forecast issuance time $t_0$, we construct supervised samples using a 72 h historical window and a 24 h forecast window at an hourly temporal resolution (i.e., 72 past steps and 24 future steps). Historical observations prior to $t_0$ are treated as reliable inputs, whereas variables after $t_0$ are represented by forecast surrogates designed to emulate operationally relevant conditions \cite{nuno2018simulation}. To avoid temporal leakage, all climatological statistics used in future feature construction are estimated from training years only. The dataset is split strictly chronologically into 2018--2022 for training, 2023 for validation, and 2024 for testing. This design is intended to support controlled comparative evaluation under a fixed temporal split rather than full climatological generalization across multiple independent test years. By construction, each of the five climate zones contributes 20 sites under this fixed split, yielding 36{,}440 training, 7{,}300 validation, and 7{,}280 test hourly issue-time samples per zone; within the 2024 test year, the retained daylight issue-hour cases break down seasonally into 1{,}780 (DJF), 1{,}840 (MAM), 1{,}840 (JJA), and 1{,}820 (SON) samples, identically across all zones.

To generate the virtual PV power, the physical conversion from meteorological inputs to AC power was simulated using pvlib. The virtual plant was configured with a generic fixed tilt system facing south (tilt angle equal to local latitude, azimuth of $180^\circ$), a nominal DC capacity of 1 kW, and an inverter efficiency of 96\%. The cell temperature was estimated using the Sandia Array Performance Model (SAPM) for an open-rack glass-polymer configuration. Furthermore, the theoretical clear-sky irradiance $\mathrm{GHI}_{\mathrm{cs}}(t)$ was calculated via the Ineichen model, with the Linke turbidity factor dynamically retrieved from the built-in climatological lookup table provided by the pvlib framework.

\subsection{Construction of Future Forecast Features}

Rather than treating all future variables as either perfectly known or uniformly perturbed, we construct future inputs according to their physical predictability. Deterministic variables, such as solar geometry and cyclic time encodings, are computed directly from astronomical relations and timestamps. Slowly varying variables, including surface albedo and aerosol related terms, are represented using persistence. For less critical drivers with strong diurnal and seasonal structure, such as dew point and relative humidity, we use diurnal and seasonal climatological expectations derived strictly from the training set, which preserves their typical temporal pattern while avoiding future information leakage.

Only the key drivers carrying substantial uncertainty in the future window are explicitly perturbed, namely GHI and ambient temperature. DNI and DHI are not independently noised. Instead, they are reconstructed from perturbed GHI through Erbs decomposition so that the future radiation field remains physically consistent \cite{erbs1982estimation,mabasa2026development}. This design keeps the perturbation mechanism focused on the principal forecast error channels while avoiding unnecessary distortion of variables that would be deterministic or weakly varying in practice.

\subsection{Dynamic NWP Error Injection Framework}

To emulate HRRR-calibrated forecast uncertainty, we inject parameterized NWP errors into future GHI and temperature using a framework calibrated against historical HRRR forecast and observation pairs \cite{lee2024importance,benjamin2025excessive} matched with NREL sites during the testing period (2024). Directly using archived HRRR forecasts would provide one historically fixed realization of forecast error for each site and period, but would not support systematic stress testing across controlled uncertainty regimes. We therefore extract the key residual statistics of HRRR errors, including heteroscedasticity related to sky state and short-term autocorrelation, and embed them into a parameterized perturbation model. This design allows the global noise strength ($s$) and temporal persistence ($\rho$) to be varied continuously, enabling the robustness boundaries of different forecasting models to be evaluated under progressively more adverse but empirically anchored conditions. The formulation combines three key elements: heteroscedasticity related to sky state, temporal persistence, and reconstruction that preserves radiation consistency.

For GHI, the error amplitude is modulated by the clearness index \cite{woyte2007fluctuations,yang2012hourly}
\begin{equation}
K_t = \min \left( \max \left( \frac{\mathrm{GHI}_{\mathrm{true}}(t)}{\mathrm{GHI}_{\mathrm{cs}}(t)}, 0 \right), 1.2 \right),
\end{equation}
where $\mathrm{GHI}_{\mathrm{cs}}(t)$ is the clear-sky global horizontal irradiance computed using the Ineichen clear-sky model \cite{ineichen2002new}. The upper bound of 1.2 allows physically plausible over-irradiance events associated with cloud enhancement, while preventing unrealistically large $K_t$ values from dominating the perturbation scale. The corresponding dynamic noise scale is
\begin{equation}
\sigma_m(t) = s \cdot \left( 1 + \alpha \cdot (1 - K_t) \right),
\end{equation}
where $s$ controls the global noise strength \cite{yagli2020reconciling}, while $\alpha$ controls the heteroscedastic amplification associated with sky state. Calibrated from HRRR residual statistics, this term enables the injected error variance to vary with the local clearness regime, thereby making the perturbation process more physically grounded.

Temporal persistence is introduced through a first-order autoregressive process
\begin{equation}
\epsilon(t) = \rho \epsilon(t-1) + \eta(t), \qquad \eta(t) \sim \mathcal{N}(0, 1-\rho^2),
\end{equation}
where $\rho \in [0,1)$ controls temporal autocorrelation while preserving unit marginal variance.

The perturbed GHI is then generated additively with respect to the clear-sky baseline,
\begin{equation}
\mathrm{GHI}_{\mathrm{noisy}}(t) = \max \left( 0,\,
\mathrm{GHI}_{\mathrm{true}}(t) + b_{\mathrm{ghi}}\mathrm{GHI}_{\mathrm{cs}}(t)
+ \mathrm{GHI}_{\mathrm{cs}}(t)\epsilon(t)\sigma_m(t) \right),
\end{equation}
which avoids nonphysical suppression under low-irradiance conditions. DNI and DHI are then reconstructed from $\mathrm{GHI}_{\mathrm{noisy}}$ using Erbs decomposition to maintain radiation consistency.

For temperature, we apply a simpler additive perturbation,
\begin{equation}
T_{\mathrm{noisy}}(t) = T_{\mathrm{true}}(t) + b_{\mathrm{temp}} + \xi(t), \qquad \xi(t) \sim \mathcal{N}(0,\sigma_{\mathrm{temp}}^2),
\end{equation}
which reflects its smoother variability relative to irradiance. Unlike solar irradiance, which is highly sensitive to rapid, localized cloud transients, ambient temperature possesses greater thermal inertia and stronger spatial continuity. This physical distinction justifies a simpler additive perturbation for temperature. Together, these components yield a perturbation framework that is statistically flexible, physically constrained, and sufficiently compact for systematic robustness experiments.

The perturbation model should nonetheless be interpreted as an HRRR-calibrated approximation rather than a complete generative model of operational NWP uncertainty. In particular, the current formulation emphasizes marginal variance modulation, short-range temporal persistence, and reconstruction that preserves radiation consistency, but it does not explicitly reproduce full spatial correlation structure, regime switching nonstationarity, or all possible dependencies among forecast errors across variables. Accordingly, the conclusions drawn in this study are intended to characterize robustness behavior within this physically constrained perturbation family, not under every possible realization of forecast uncertainty.

\subsection{Benchmark Models and Training Setup}

To prevent the accumulation of iterative errors over the extended 24-h forecast horizon, all evaluated models employ a direct multi-step (DMS) forecasting strategy. For the deep learning architectures, the networks are designed in a Multi-Input Multi-Output (MIMO) configuration, directly mapping the 72-h historical context and future NWP inputs to a 24-dimensional continuous power output vector. For LightGBM, 24 independent regressors are trained, one for each specific lead time. In this study, LightGBM is treated as a strong tabular baseline rather than as a lightweight reference, because it uses engineered predictors specific to each forecast horizon derived from the same underlying information set as the sequence models.

Although the deep learning and tree-based models share the same raw information sources, their input representations differ. The deep learning models operate directly on the full spatiotemporal tensor composed of historical power, future meteorological drivers, deterministic time encodings, and auxiliary physical variables. By contrast, LightGBM converts this tensor into tabular features specific to each forecast horizon, including the target step future predictors, the corresponding clearness index, and multi-scale summaries of the historical power sequence (3 h, 6 h, and 24 h rolling means and standard deviations; 1 h, 2 h, and 24 h lags; and the 1 h first-order difference). The resulting feature usage is summarized in Table~\ref{tab:input_features}.

Accordingly, the comparison in this study should not be interpreted as a pure architecture only ablation under identical feature representation. Rather, it evaluates practically relevant model families under matched underlying information sources but representations appropriate to each model type. The intent is to compare robustness behavior under representative usage patterns for sequence and tabular models, while making the representation asymmetry explicit.

\begin{table}[htbp]
    \centering
    \caption{Shared input information and representations for each model type. Both model families use the same underlying forecast information, but sequence models consume the full tensor directly, whereas LightGBM uses tabular features constructed for each forecast horizon.}
    \label{tab:input_features}
    \small
    \setlength{\tabcolsep}{6pt}
    \begin{tabular}{>{\centering\arraybackslash}p{2.8cm} @{\hspace{1.5em}} >{\centering\arraybackslash}p{4.2cm} c c p{4.6cm}}
        \toprule
        Feature group & Example variables & DL & LGBM & Representation \\
        \midrule
        Historical target dynamics & Past PV power (72-h window) & Yes & Yes & DL: Full sequence \newline LGBM: 1, 2, 24 h lags; 3, 6, 24 h rolling stats; 1 h diff. \\
        \addlinespace
        Future irradiance drivers & Future GHI and reconstructed radiation channels & Yes & Yes & DL: Future sequence \newline LGBM: Target-horizon predictors \\
        \addlinespace
        Future temperature drivers & Future ambient temp. \& related thermodynamics & Yes & Yes & DL: Future sequence \newline LGBM: Target-horizon predictors \\
        \addlinespace
        Deterministic time encodings & Hour-of-day, day-of-year cyclic encodings & Yes & Yes & DL: Sequence channels \newline LGBM: Target-step covariates \\
        \addlinespace
        Physical priors \& aux. meteorology & Solar geometry, clear-sky quantities, aux. variables & Yes & Yes & DL: Sequence channels \newline LGBM: Target-step covariates \\
        \addlinespace
        Explicit engineered summaries & Clearness index $K_t$, recent stats, $\Delta$power (1~h) & Implicit & Yes & Constructed explicitly for each forecast horizon \\
        \bottomrule
    \end{tabular}
\end{table}

Prior to training, the deep learning inputs are standardized via Z-score normalization. Crucially, to strictly prevent temporal information leakage, the normalization statistics (mean and standard deviation) are computed exclusively from the historical observation window within the training set, explicitly excluding any future forecast variables from the scaling calculations.
\begin{equation}
\tilde{x} = \frac{x-\mu_{\mathrm{train,hist}}}{\sigma_{\mathrm{train,hist}}}.
\end{equation}
Here, $x$ denotes an input feature value in the deep learning tensor, while $\mu_{\mathrm{train,hist}}$ and $\sigma_{\mathrm{train,hist}}$ denote the corresponding mean and standard deviation estimated only from the historical portions of the training samples.

For evaluation, let $y_i$ and $\hat{y}_i$ denote the ground truth and predicted PV power for sample $i$, respectively. The normalized mean absolute error (NMAE) and normalized root mean square error (NRMSE) are defined as
\begin{equation}
\mathrm{NMAE}=\frac{\frac{1}{N}\sum_{i=1}^{N}\left|\hat{y}_i-y_i\right|}{\frac{1}{N}\sum_{i=1}^{N}\left|y_i\right|},
\end{equation}
\begin{equation}
\mathrm{NRMSE}=\frac{\sqrt{\frac{1}{N}\sum_{i=1}^{N}\left(\hat{y}_i-y_i\right)^2}}{\frac{1}{N}\sum_{i=1}^{N}\left|y_i\right|}.
\end{equation}
In the lead-time robustness analysis, we further use a clear-sky-normalized RMSE, hereafter referred to as Relative Error (RE). This metric is used specifically for lead-time diagnostics because PV power has a strong diurnal cycle: without clear-sky normalization, absolute errors around midday would be naturally larger than those near sunrise or sunset simply because the available solar resource is higher. Normalizing by the clear-sky PV power profile reduces the influence of deterministic solar-geometry-driven intra-day variation, allowing the analysis to focus more directly on how algorithmic forecast errors propagate and accumulate with increasing lead time.
\begin{equation}
\mathrm{RE}=\frac{\sqrt{\frac{1}{N}\sum_{i=1}^{N}\left(\hat{y}_i-y_i\right)^2}}{\frac{1}{N}\sum_{i=1}^{N}\max\left(P_{\mathrm{cs},i},\varepsilon\right)},
\end{equation}
where $P_{\mathrm{cs},i}$ is the clear-sky PV power corresponding to sample $i$, and $\varepsilon=10^{-6}$ is a small positive numerical floor introduced only to avoid division by zero.

The deep learning models are optimized using the AdamW optimizer with a learning rate of $1 \times 10^{-3}$, a weight decay of $1 \times 10^{-4}$, and a batch size of 512. The objective function is the Mean Squared Error (MSE). To ensure training stability, gradient clipping is applied with a maximum norm of 1.0. The training process runs for a maximum of 40 epochs, with early stopping applied if the validation loss fails to improve for 8 consecutive epochs. Similarly, the LightGBM models are configured with 500 estimators and an early stopping round of 30 based on the L2 validation metric. This setup is intended to provide a deliberately strong tabular baseline under the same DMS forecasting task, rather than a minimal tree-based reference.

For reproducibility, all main experiments use a fixed random seed of 42. Hyperparameters were selected as stable shared settings across climate zones and perturbation scenarios rather than exhaustively retuned for each individual case, so that the comparison emphasizes robustness trends under a common protocol. Training and attribution experiments were conducted on a server equipped with an NVIDIA Quadro RTX 8000 GPU. Unless otherwise noted, the latency results in Section 6 were measured on CPU in end-to-end single-sample mode for the full 24-step horizon, including any required feature construction.

\section{Baseline Performance and Robustness Evaluation}

To comprehensively quantify the performance boundaries of different models under complex NWP forecast errors, this section first establishes baselines under clean, no noise conditions, and then evaluates how forecast errors evolve with noise magnitude, temporal autocorrelation, and forecast lead time.

\subsection{Baseline Forecasting Performance under Clean Data}

\begin{figure}[H]
    \centering
    \includegraphics[width=0.90\textwidth]{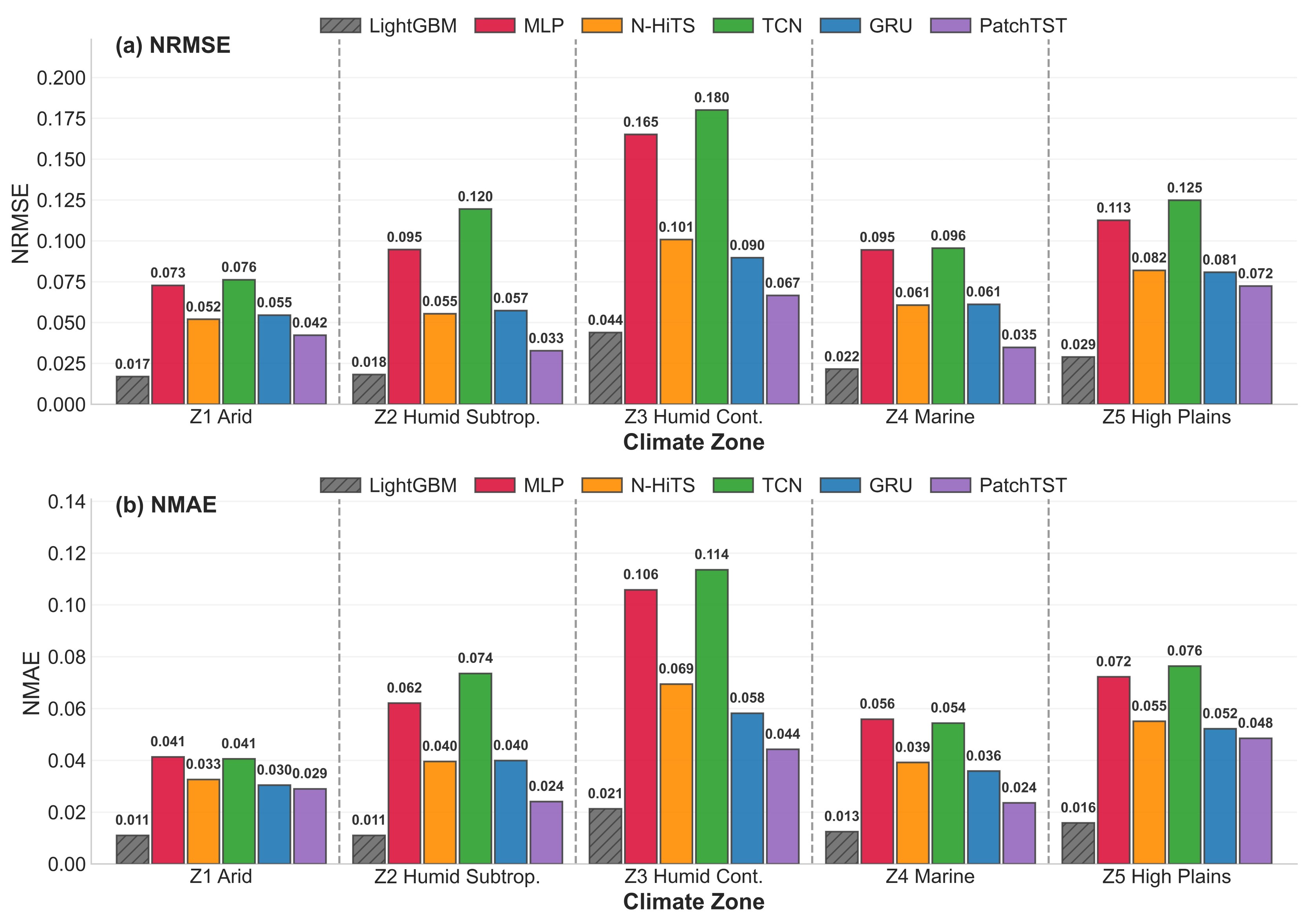}
    \caption{Baseline forecasting performance of the six evaluated models using clean (unperturbed) input data. Performance is quantified by (a) normalized root mean square error (NRMSE) and (b) normalized mean absolute error (NMAE) across the five designated climate zones.}
    \label{fig:fig2}
\end{figure}

We first establish the baseline forecasting capabilities of the six models using idealized, unperturbed inputs. As illustrated in Fig.~\ref{fig:fig2}, the models exhibit a consistent performance hierarchy across all five climate zones. Under perfect forecast conditions, LightGBM demonstrates a clear advantage in this tabular forecasting task, which is consistent with its role as a strong engineered tabular baseline. Its NRMSE values remain below 0.03 in four of the five climate zones, with Z3 Humid Continental standing out as the most challenging region. Among the deep learning architectures, PatchTST and GRU consistently deliver the most competitive performance. Conversely, Multi-Layer Perceptron (MLP) \cite{bai2018empirical} and Temporal Convolutional Network (TCN) \cite{zeng2023transformers} yield substantially higher errors, particularly in meteorologically complex regions such as Z3.

\subsection{Non-linear Impact of Noise Magnitude and Auto-correlation}

\begin{figure}[H]
    \centering
    \includegraphics[width=0.98\textwidth]{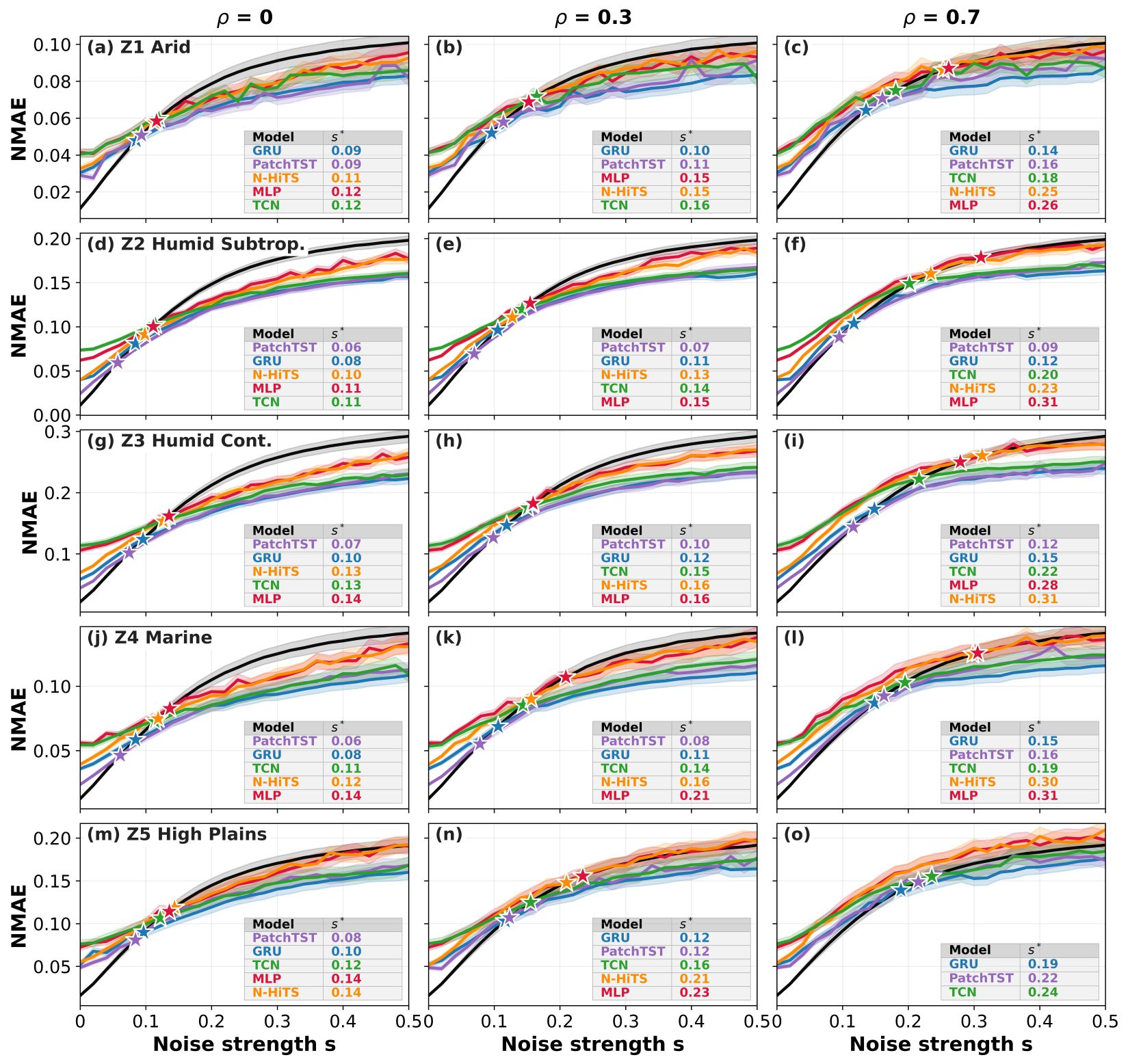}
    \caption{Forecasting error progression as a function of injected noise strength $s$. The subplots detail the NMAE response under three distinct temporal autocorrelation levels of the injected noise ($\rho = 0$, $0.3$, and $0.7$). Rows correspond to the five climate zones (Z1--Z5). Embedded tables within each subplot summarize the relative model rankings based on the critical noise threshold ($s^*$).}
    \label{fig:fig3}
\end{figure}

To quantitatively capture this performance shift, we define a critical noise threshold $s^*$, representing the specific injected noise magnitude $s$ at which a deep learning model begins to outperform the LightGBM baseline. A lower $s^\ast$ indicates stronger robustness relative to the tabular baseline, because the model overtakes LightGBM at an earlier disturbance level. The evaluation in Fig.~\ref{fig:fig3} reveals two distinct regimes of model superiority. Under low noise conditions, LightGBM functions as an exceptionally strong baseline due to its efficient utilization of engineered tabular features. However, its performance degrades rapidly as the noise magnitude $s$ increases, particularly in zones with variable cloud conditions (Z2, Z3, Z5). In contrast, sequence models such as GRU and PatchTST exhibit much flatter degradation curves, suggesting stronger noise filtering capability in medium to high noise environments. The critical threshold $s^*$ is highly sensitive to the temporal structure of errors. Under white noise conditions ($\rho=0$), the crossover occurs at very low $s^*$ (e.g., $s^* \approx 0.06$ for PatchTST in Z2), indicating immediate deep learning superiority under random perturbations. However, as error persistence increases ($\rho=0.7$), these thresholds shift upward substantially, as seen in Z5 where $s^*$ for GRU rises to 0.19. Notably, in certain extreme scenarios such as Z5 with high $\rho$ (Fig. 3o), simpler architectures such as MLP and N-HITS fail to cross the LightGBM baseline within the evaluated range. This absence of a crossover point underscores the limited robustness of nonsequential or shallow architectures under highly autocorrelated NWP errors.

\begin{figure}[H]
    \centering
    \includegraphics[width=0.95\textwidth]{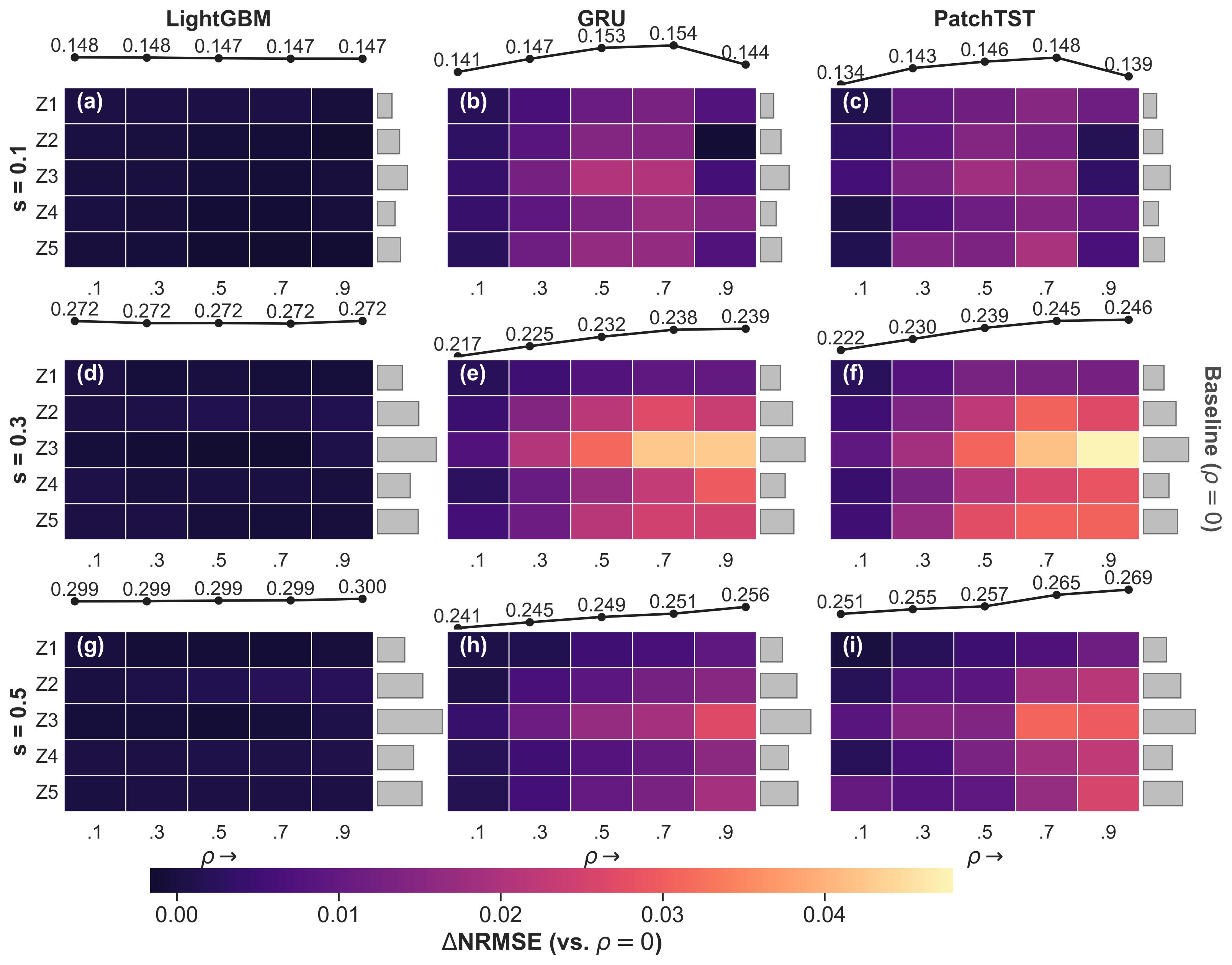}
    \caption{Heatmaps illustrating the performance degradation of LightGBM, GRU, and PatchTST relative to the baseline ($\Delta$NRMSE), where $\Delta$NRMSE is computed as $\mathrm{NRMSE}(s,\rho)-\mathrm{NRMSE}(s,\rho=0)$ across varying noise magnitudes and temporal autocorrelation levels.}
    \label{fig:fig4}
\end{figure}

Figure~\ref{fig:fig4} further disentangles these coupled effects through heatmaps of performance degradation relative to the baseline ($\Delta$NRMSE), defined as
\begin{equation}
\Delta \mathrm{NRMSE}(s,\rho)=\mathrm{NRMSE}(s,\rho)-\mathrm{NRMSE}(s,\rho=0).
\end{equation}

LightGBM appears remarkably insensitive to the temporal structure of the errors. Once $s$ is fixed, its error remains nearly unchanged across the entire $\rho$ spectrum. For example, its mean NRMSE stays around 0.272 at $s=0.3$ and changes only marginally from 0.299 to 0.300 at $s=0.5$ as $\rho$ increases. This result indicates that the primary vulnerability of tree-based models lies in the absolute magnitude of the input perturbation, as they process each time step independently.

In contrast, the deep sequence models (GRU and PatchTST) exhibit a pronounced vulnerability to error persistence. As $\rho$ approaches 0.9—representing an extreme stress test scenario that exceeds typical HRRR error persistence (average $\rho \approx 0.5$)—the performance of even the most robust sequence models begins to degrade significantly. This compounding effect becomes particularly severe under elevated noise magnitudes and is geographically amplified in meteorologically challenging regions, such as the Humid Continental zone (Z3). These distinct degradation trajectories reveal a fundamental architectural trade-off: while deep time series models effectively filter out independent random shocks, highly autocorrelated errors act as sustained disturbances. Such persistent deviations are more likely to disrupt their internal temporal attention or gated memory mechanisms, ultimately inducing a cascaded bias over the extended forecast horizon.

\subsection{Error Propagation and Accumulation across Forecast Horizons}

\begin{figure}[htbp!]
    \centering
    \includegraphics[width=0.98\textwidth, height=0.85\textheight, keepaspectratio]{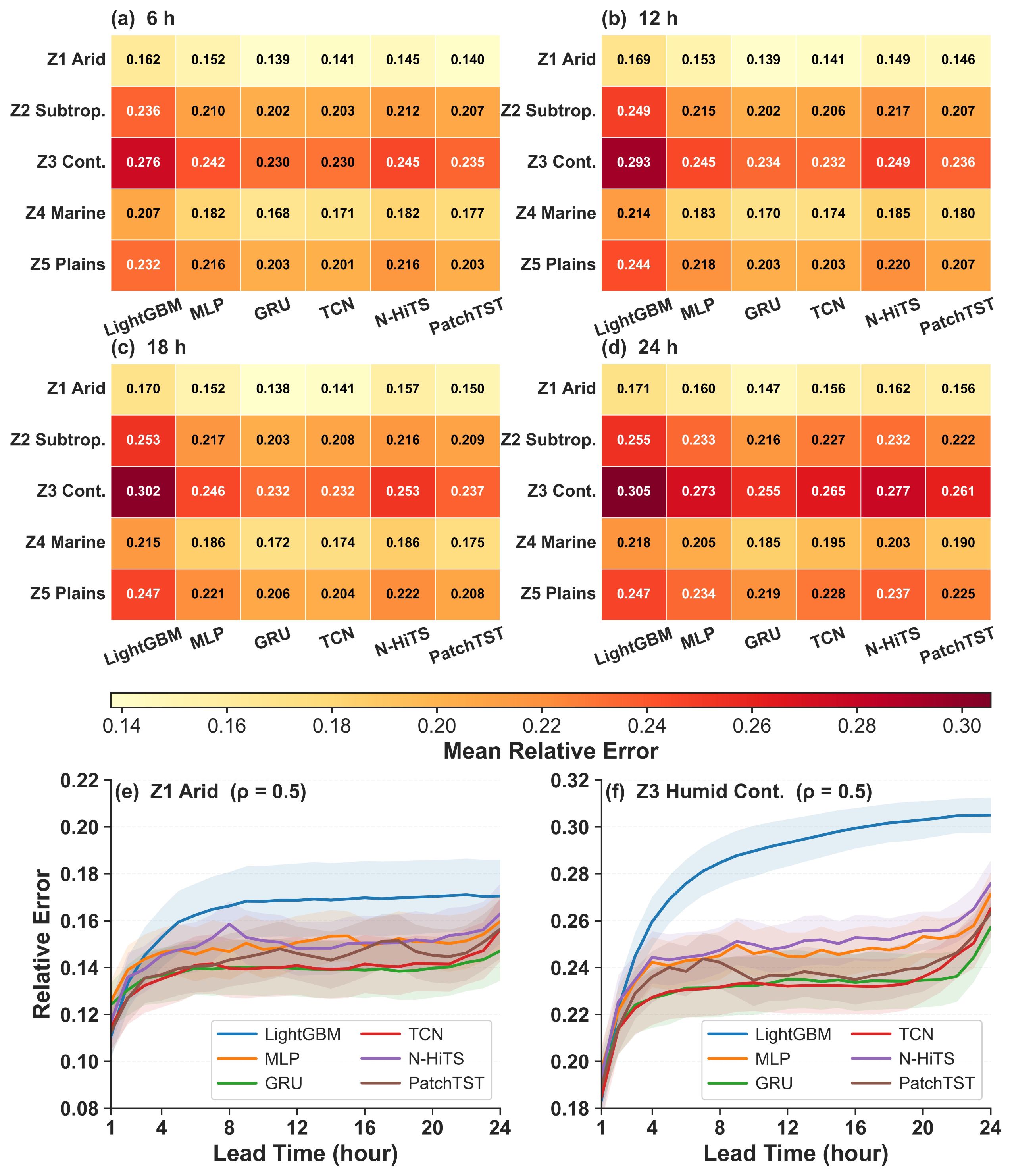}
    \caption{Evaluation of forecasting performance across different forecast lead times. (a)--(d) Heatmaps illustrating the mean Relative Error (i.e., clear-sky-normalized RMSE) of the six evaluated models across the five climate zones at specific lead times of 6, 12, 18, and 24 hours. (e)--(f) Temporal progression of the Relative Error over a continuous 24 h lead time horizon for the Z1 Arid and Z3 Humid Continental zones, evaluated under an error autocorrelation condition of $\rho=0.5$.}
    \label{fig:fig5}
\end{figure}

The use of Relative Error (RE) allows for a focused analysis of forecasting deviations by normalizing the absolute error against the clear-sky irradiance profile. NWP errors do not act as isolated temporal perturbations. Rather, their impact propagates and accumulates continuously throughout the forecast horizon \cite{wang2022cost,yang2022correlogram}. Figure~\ref{fig:fig5} quantifies this temporal cascading effect, detailing the evolution of the mean RE across the multi-horizon forecast window up to 24 hours. A universal trend across all climate zones is the progressive amplification of forecasting errors as the lead time extends. However, the rate of this degradation exhibits spatial heterogeneity. The Humid Continental zone (Z3), characterized by frequent and complex cloud transitions, consistently exhibits the most rapid error accumulation, reaching average RE values of roughly 0.255--0.305 at the 24 h lead time. In contrast, the Arid zone (Z1) experiences much slower error propagation, with 24 h errors remaining in a narrower range of about 0.147--0.171 across models.

From an architectural perspective, the temporal propagation heatmaps underscore the vulnerability of static tabular models over extended horizons. While LightGBM is often favored for its tabular efficiency, it exhibits immediate and rapid error accumulation. Its predictive advantage over sequence models vanishes within the first few lead time steps, ultimately yielding the highest 24 h error across all climate zones. The temporal cascading effect varies starkly across meteorological regimes. In the stable Arid zone (Z1), error propagation remains constrained across all models. However, in the cloud dynamic Z3 region, the lack of recurrent memory in LightGBM leads to a sharp escalation in RE, peaking at 0.305. By contrast, deep sequence models, particularly GRU, exhibit stronger temporal resilience, with substantially flatter error growth trajectories over the forecast horizon. Across all five climate zones, GRU consistently maintains the lowest 24 h forecasting error, closely followed by PatchTST, highlighting the necessity of temporal dependency modeling for reliable long horizon PV predictions.

Figures~\ref{fig:fig5}(e) and \ref{fig:fig5}(f) further indicate that error growth is concentrated mainly in the first 4--8 hours and then transitions into a more gradual plateau. This behavior is especially pronounced in Z3, where LightGBM rises rapidly from roughly 0.19 to above 0.29, whereas GRU and PatchTST maintain markedly flatter trajectories across the full 24 h horizon. By contrast, the curves in Z1 remain substantially flatter, indicating weaker cascading amplification under more stable radiation backgrounds. The shaded bands in Figs.~\ref{fig:fig5}(e) and \ref{fig:fig5}(f) denote the $\pm 1$ standard deviation across sites within the same climate zone. They therefore reflect spatial dispersion within each zone rather than confidence intervals of the mean.

\FloatBarrier

\section{Physical Drivers and Error Attribution}

After quantifying the macroscopic robustness boundaries of the models under NWP errors, this section further explains these degradation phenomena from the dual perspectives of empirical error statistics and physical climate mechanisms.

\subsection{Empirical Residual Structure of HRRR and Parameter Calibration}

\begin{figure}[htbp!]
    \centering
    \includegraphics[width=0.98\textwidth, height=0.85\textheight, keepaspectratio]{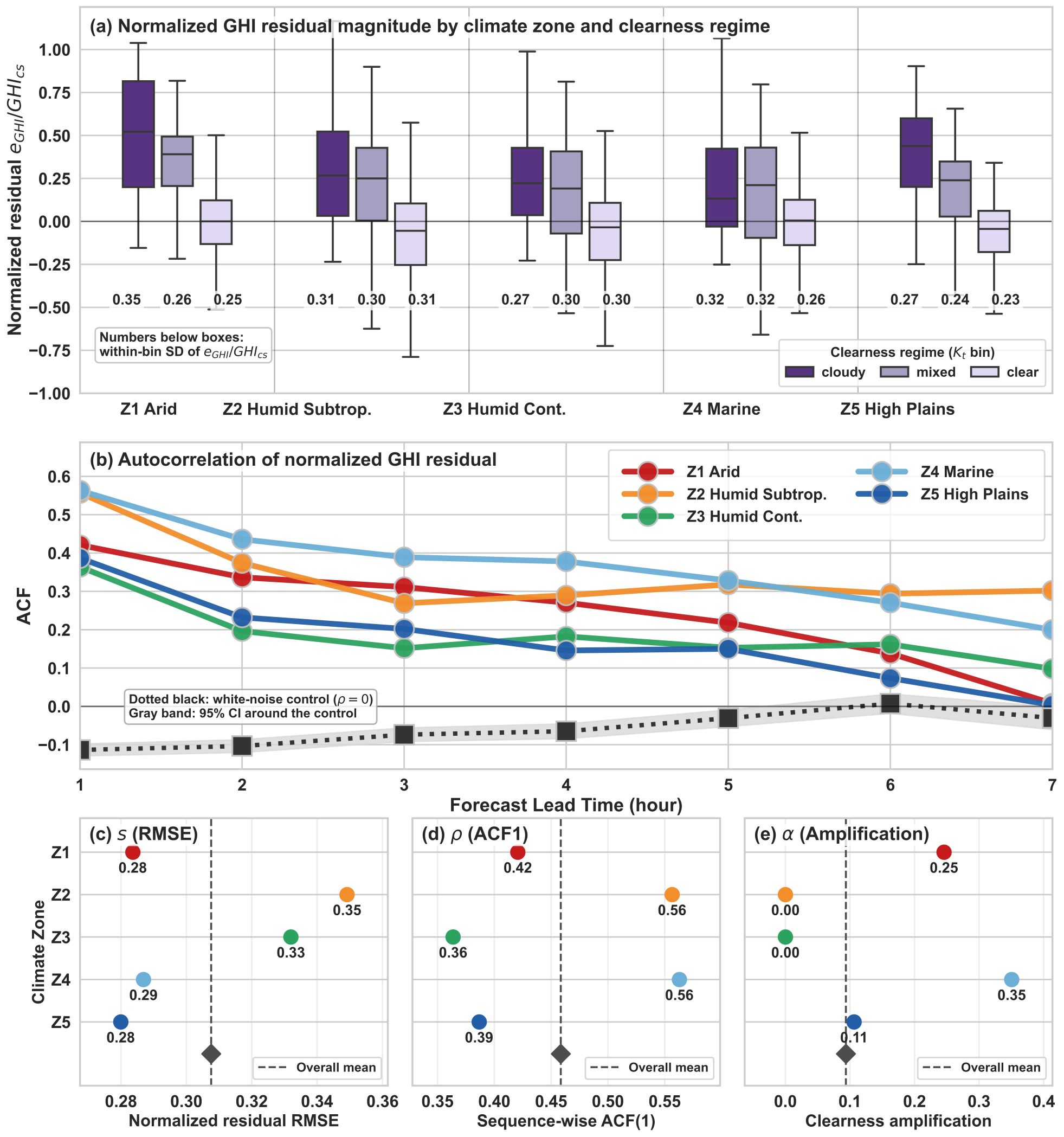}
    \caption{Statistical characteristics of the GHI prediction residuals. The normalized residual is defined as $e_{\mathrm{GHI}}^{\mathrm{norm}}=(\mathrm{GHI}_{\mathrm{HRRR}}-\mathrm{GHI}_{\mathrm{obs}})/\max(\mathrm{GHI}_{\mathrm{cs}},\varepsilon)$ and is evaluated only for daytime samples with solar elevation $\ge 10^\circ$ and $\mathrm{GHI}_{\mathrm{cs}} \ge 100$ W m$^{-2}$. The clearness regimes used for residual analysis are defined as cloudy ($K_t^{\mathrm{obs}} \le 0.3$), mixed ($0.3 < K_t^{\mathrm{obs}} \le 0.7$), and clear ($K_t^{\mathrm{obs}} > 0.7$). (a) Distribution of normalized GHI residual magnitude categorized by climate zone and clearness regime. (b) Temporal autocorrelation function (ACF) of the normalized GHI residuals over forecast lead times. (c)--(e) Spatial distribution of extracted error injection parameters: baseline RMSE ($s$), ACF at lag 1 estimated along each sequence ($\rho$), and clearness amplification factor ($\alpha$).}
    \label{fig:fig6}
\end{figure}

NWP forecast errors are not uniformly distributed but are strongly modulated by local sky conditions and regional climatology. Figure~\ref{fig:fig6} characterizes the empirical statistical structure of normalized HRRR GHI residuals, directly validating the parameterizations used in our dynamic error injection framework. Specifically, the normalized irradiance residual is defined as
\begin{equation}
e_{\mathrm{GHI}}^{\mathrm{norm}}(t)=\frac{\mathrm{GHI}_{\mathrm{HRRR}}(t)-\mathrm{GHI}_{\mathrm{obs}}(t)}{\max\left(\mathrm{GHI}_{\mathrm{cs}}(t),\varepsilon\right)},
\end{equation}
and the recommended noise magnitude and persistence parameters are calibrated from these residuals as
\begin{equation}
s=\sqrt{\frac{1}{N}\sum_{i=1}^{N}\left(e_{\mathrm{GHI},i}^{\mathrm{norm}}\right)^2}, \qquad
\rho=\mathrm{Corr}\!\left(e_{\mathrm{GHI}}^{\mathrm{norm}}(t),e_{\mathrm{GHI}}^{\mathrm{norm}}(t-1)\right),
\end{equation}
where $\rho$ is computed along each forecast trajectory and then aggregated by climate zone. To avoid unstable normalization during twilight or under very weak clear-sky irradiance, only daytime samples satisfying solar elevation $\ge 10^\circ$ and $\mathrm{GHI}_{\mathrm{cs}} \ge 100$ W m$^{-2}$ are retained in the normalized GHI residual analysis. Accordingly, the reported bias, MAE, RMSE, standard deviation, and interquartile range in Fig.~\ref{fig:fig6} are all calculated from this filtered residual set, while the lag-1 autocorrelation is estimated on each sequence before aggregation.

The heteroscedastic amplification parameter $\alpha$ is estimated from summaries by clearness bin rather than from individual samples. Specifically, for the residual calibration analysis in Fig.~\ref{fig:fig6}, the filtered residuals are grouped into three observed clearness regimes defined as cloudy ($K_t^{\mathrm{obs}} \le 0.3$), mixed ($0.3 < K_t^{\mathrm{obs}} \le 0.7$), and clear ($K_t^{\mathrm{obs}} > 0.7$). For each bin $b$, we compute the standard deviation $\sigma_b$ of $e_{\mathrm{GHI}}^{\mathrm{norm}}$ together with the corresponding bin midpoint $K_b$. Taking the clearest bin as the reference with midpoint $K_{\mathrm{ref}}$ and standard deviation $\sigma_{\mathrm{ref}}$, we fit
\begin{equation}
\frac{\sigma_b}{\sigma_{\mathrm{ref}}}-1 \approx \alpha \, \max\!\left(K_{\mathrm{ref}}-K_b,0\right),
\end{equation}
using the bin sample counts as regression weights. The fitted slope is constrained to be nonnegative and then used as the recommended $\alpha$ in the continuous noise scale model of Eq.~(3). In this way, $\alpha$ quantifies how rapidly the residual spread broadens as sky conditions depart from the clearest regime, and it measures how much operational forecast uncertainty expands during active cloud transient conditions relative to clear-sky conditions.

Residual distributions under cloudy and mixed skies are substantially wider than those under clear skies. This heteroscedastic behavior provides strong empirical justification for the clearness based modulation factor $\alpha$ employed in this study. Furthermore, the residuals exhibit significant temporal memory. Unlike the white noise control group, all five climate zones display pronounced positive autocorrelation at short-term lags, indicating that real NWP errors act as persistent disturbances rather than independent noise. Specifically, the lag-1 autocorrelation (ACF(1)) estimated along each sequence is most pronounced in the Humid Subtropical (Z2) and Marine (Z4) zones, indicating that forecast errors in these moist environments possess stronger temporal memory. Furthermore, regional parameter estimates reveal that the largest normalized residual strengths ($s$) correspond to Z2 and Z3, while heteroscedastic amplification ($\alpha$) peaks in Z4, underscoring how specific local climates uniquely shape the empirical error structure. Overall, the spatial distributions of $s$, $\rho$, and $\alpha$ extracted from these HRRR statistics indicate that our simulated error environments are anchored to empirical atmospheric dynamics \cite{mathiesen2011evaluation,jimenez2016wrf}.

\FloatBarrier

\subsection{Meteorological Physical Drivers of Error Amplification}

\begin{figure}[H]
    \centering
    \includegraphics[width=0.95\textwidth]{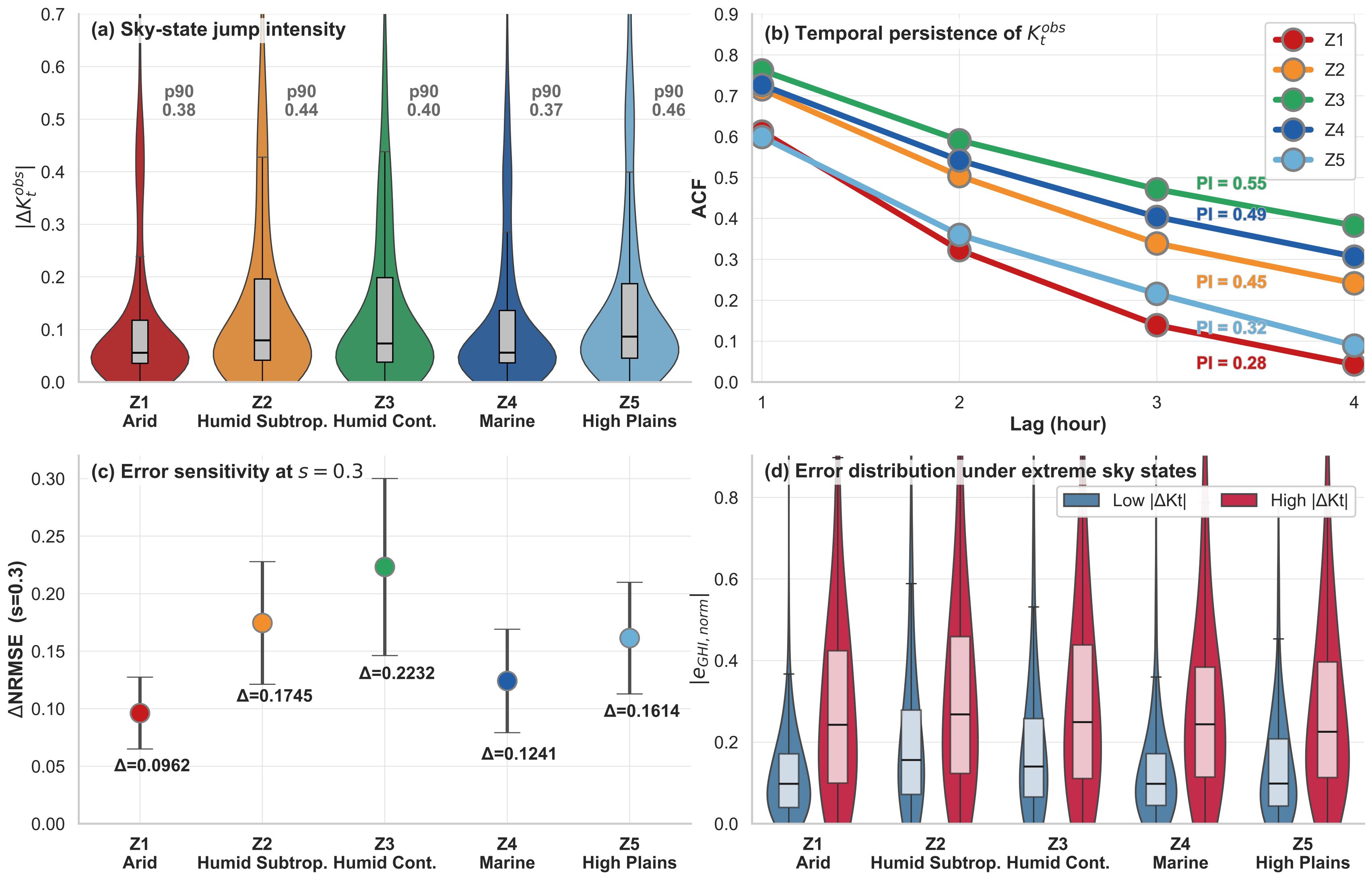}
    \caption{Physical meteorological drivers of forecasting errors and their spatial heterogeneity. (a) Distribution of sky state jump intensity across the climate zones. (b) Temporal persistence of the observed clearness index across varying time lags. (c) Average performance degradation by zone, measured as $\Delta$NRMSE at the fixed noise strength $s=0.3$ relative to the clean baseline. (d) Distribution of normalized errors under low and high jump regimes, where the two regimes are defined within each climate zone using the zone specific median of $|\Delta K_t^{\mathrm{obs}}|$ as the threshold.}
    \label{fig:fig7}
\end{figure}

While empirical residuals define the boundary conditions of input perturbations, they do not fully explain why identical error magnitudes induce substantially different levels of forecasting degradation across regions. Figure~\ref{fig:fig7} bridges this gap by attributing the spatial heterogeneity of model robustness to local sky state evolution.

We introduce sky state jump intensity $|\Delta K_t^{\mathrm{obs}}|$ to quantify the magnitude of short-term radiation background fluctuations, where the observed clearness index is
\begin{equation}
K_t^{\mathrm{obs}}(t)=\min\left(\max\left(\frac{\mathrm{GHI}_{\mathrm{obs}}(t)}{\max\left(\mathrm{GHI}_{\mathrm{cs}}(t),\varepsilon\right)},0\right),1.2\right),
\end{equation}
and the jump intensity is
\begin{equation}
\left|\Delta K_t^{\mathrm{obs}}(t)\right|=\left|K_t^{\mathrm{obs}}(t)-K_t^{\mathrm{obs}}(t-1)\right|.
\end{equation}
To summarize short-term structural memory, we define a persistence index as the mean lagged autocorrelation of $K_t^{\mathrm{obs}}$ over the first four lags,
\begin{equation}
\mathrm{PI}=\frac{1}{4}\sum_{\ell=1}^{4}\mathrm{ACF}_{K_t^{\mathrm{obs}}}(\ell).
\end{equation}
Regions such as the High Plains (Z5) and Humid Subtropical (Z2) experience the most intense sky state transitions, reflecting frequent and rapid cloud evolution. Conversely, the Humid Continental zone (Z3) exhibits the highest short-term structural persistence. These distinct meteorological characteristics govern how input perturbations are nonlinearly amplified within the forecasting systems.

When evaluated under a fixed noise magnitude, the average performance degradation by zone in Fig.~\ref{fig:fig7}(c) follows these physical traits. Specifically, panel (c) reports the mean $\Delta$NRMSE at $s=0.3$ relative to the clean baseline, averaged across the evaluated model set. Z3 shows the highest overall sensitivity to input errors, whereas the structurally stable Arid zone (Z1) and the low jump Marine zone (Z4) remain comparatively insulated. This pattern suggests that spatial disparities in model robustness are not solely a function of input error magnitude. They are fundamentally constrained by whether local atmospheric dynamics amplify the coupling between input perturbations and the PV power response \cite{jarvela2020characteristics,rosenfeld2006aerosol,cui2017characterizing}.

Figure~\ref{fig:fig7}(d) provides consistent evidence from the error distributions themselves: across all climate zones, the high jump regime is associated with visibly broader distributions and heavier upper tails than the low jump regime, indicating that rapid cloud formation, dissipation, or advection is more likely to induce large forecast deviations. Ultimately, the spatial heterogeneity in model robustness is deeply rooted in local atmospheric dynamics. While the empirical HRRR residuals dictate the baseline error exposure for each region, it is the local sky state evolution—specifically, the intensity and persistence of cloud transitions—that largely determines how severely these errors propagate through the forecasting architectures.

\FloatBarrier

\section{Model Interpretability}

Building upon the macroscopic robustness evaluation, we leverage explainable AI \cite{machlev2022explainable,salih2025perspective} to investigate how the predictive reliance of the models redistributes under degraded forecast inputs. Figure~\ref{fig:fig8} presents a dual view attribution analysis on a representative case: SHAP values characterize global feature reallocation at the variable level, while IG attributions trace local migration in temporal dependence. To assess whether these patterns persist more broadly beyond the representative case, we further report consistency summaries at the population level in Supplementary Fig.~\ref{fig:supp_interpretability_consistency} and Supplementary Table~\ref{tab:supp_interpretability_stats}. Together, these results provide illustrative evidence for an adaptive tendency, which we refer to here as feature substitution.

\subsection{Global Feature Reallocation and Substitution Effect}

Under ideal clean forecast conditions, the model's decision logic is more direct. The future clearness index and global horizontal irradiance overwhelmingly dominate the predictions, accounting for 47.7\% and 44.7\% of the absolute SHAP share, respectively. In this regime, the model largely behaves as a direct mapping function from high quality future radiation to PV power, with solar geometry and historical statistics playing only supporting roles.

However, the introduction of high magnitude noise substantially reorders this importance hierarchy. In Fig.~\ref{fig:fig8}(c), the contribution share of Kt Forecast decreases by about 41 percentage points relative to the clean baseline, and GHI Forecast also declines sharply. The model correspondingly pivots toward three classes of alternative information: deterministic physical priors unaffected by NWP errors, historical power dynamics reflecting short-term system inertia, and less heavily perturbed auxiliary meteorological variables. Under highly autocorrelated errors, this reallocation pattern becomes more gradual and context dependent rather than fully abandoning future information. Figure~\ref{fig:fig8}(e) shows that GHI Forecast remains one of the most strongly reduced features, by roughly 10.2 percentage points, while variables such as Single Scattering Albedo (SSA), $\Delta$ Power (1h), Ozone, and Power Lag (1h) become more prominent. This suggests that feature substitution is not a simple fallback to a single variable, but a case level redistribution of predictive dependence under different error regimes.

Note that for the high autocorrelation scenario (Fig. 8e and 8f), a moderate noise magnitude ($s=0.1$) is intentionally selected. This experimental control isolates the specific structural impact of error persistence ($\rho=0.5$) on the model's temporal receptive field, preventing the attribution signals from being completely overwhelmed by extreme magnitude shocks.

\subsection{Temporal Attention Shift and Backward Reliance}

Beyond variable level weight reorganization, the temporal receptive field of deep sequence models fundamentally shifts under uncertainty. In the no noise baseline, attribution peaks are sharply concentrated in the immediate future window, indicating that the model derives its primary discriminative basis directly from the forecast sequence while using the historical window merely as secondary context.

\begin{figure}[htbp!]
    \centering
    \includegraphics[width=0.98\textwidth, height=0.85\textheight, keepaspectratio]{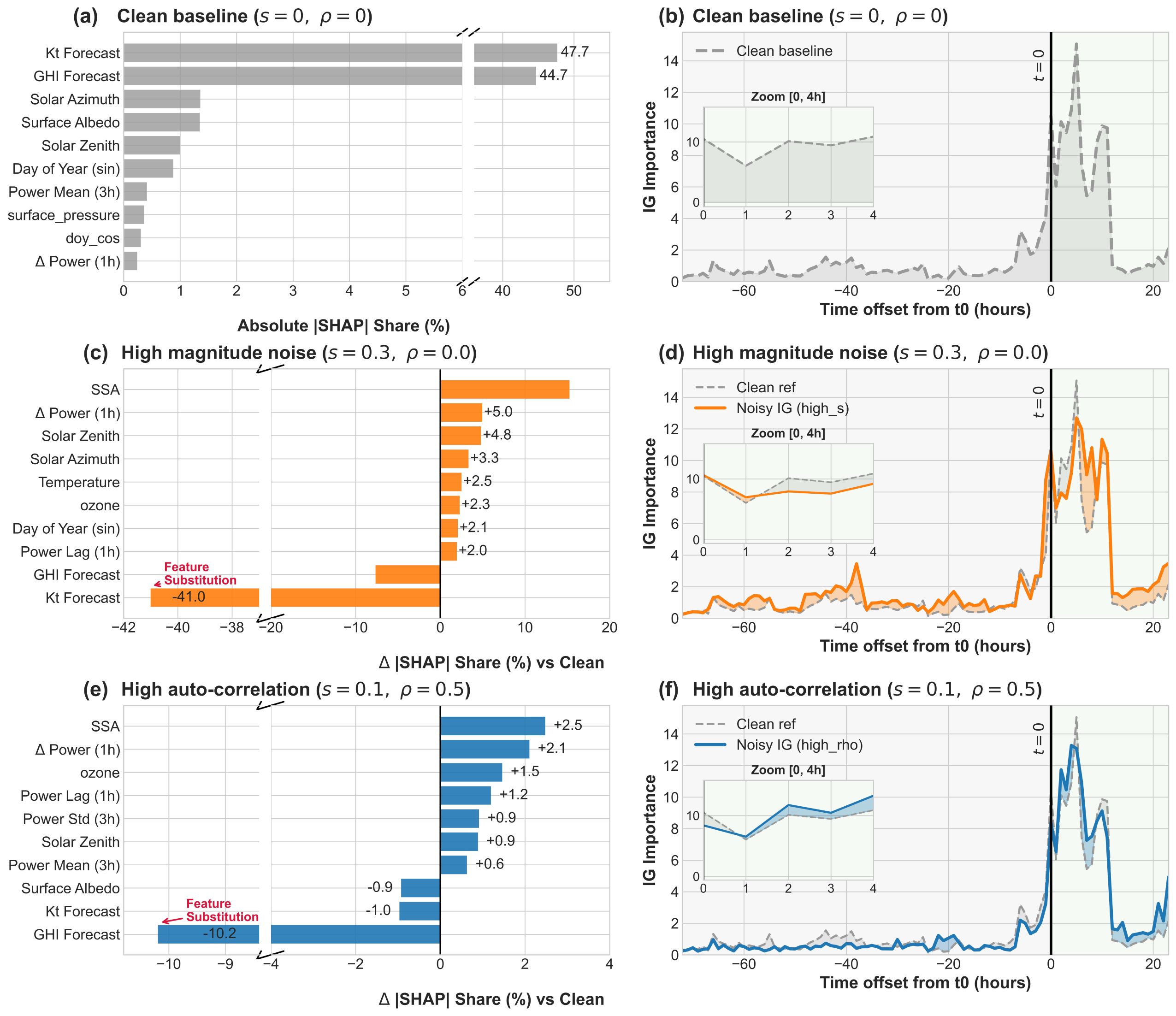}
    \caption{Model interpretability and feature attribution analysis under varying error regimes. (a), (c), and (e) Global feature importance quantified by absolute SHAP value shares for the clean baseline, and the relative change under high magnitude noise and high autocorrelation conditions, highlighting a feature substitution pattern in the representative case. (b), (d), and (f) Temporal Integrated Gradients importance over time offset, showing the model's reliance on historical versus forecast inputs.}
    \label{fig:fig8}
\end{figure}

When subjected to high magnitude noise, these sharp future peaks exhibit relative attenuation, while the attribution intensity across the historical window shows a subtle but observable elevation. This suggests a compensatory backward reliance: the model appears to supplement its compromised assessment of the future state with verified past observations. From an engineering control perspective, this behavior can be interpreted as resembling a dynamic low-pass filtering effect, in which the model draws more heavily on historical structural inertia when fast-varying future inputs become unreliable. This temporal dispersion is further exacerbated under high autocorrelation errors, where the historical interval maintains a continuous, nonzero attribution intensity to counterbalance the prolonged unreliability of the forecast sequence. Importantly, Figs.~\ref{fig:fig8}(b), \ref{fig:fig8}(d), and \ref{fig:fig8}(f) correspond to one representative case rather than an average over the full sample population, and therefore should be interpreted as illustrative evidence of this compensatory behavior, rather than a strict statistical pattern universally shared by all samples.

Supplementary Fig.~\ref{fig:supp_interpretability_consistency} and Supplementary Table~\ref{tab:supp_interpretability_stats} extend the representative case evidence in Fig.~\ref{fig:fig8} to 250 paired daytime test samples drawn from the five climate zones. Under high magnitude noise, the SHAP share of future forecast features decreases by a median of 39.1 percentage points, while the historical power, deterministic physical, and auxiliary meteorology categories increase by medians of 12.8, 5.4, and 18.3 percentage points, respectively (paired Wilcoxon signed-rank tests, all $p < 0.001$). Under high autocorrelation, the future forecast share still decreases markedly (median 10.1 percentage points), while historical power and auxiliary meteorology shares increase by medians of 4.6 and 5.0 percentage points, respectively (both $p < 0.001$). The deterministic physical category does not show a significant shift in the high autocorrelation setting ($p = 0.224$), indicating that the reallocation pathway is recurrent but not fully uniform across perturbation types. Consistent with this pattern, the historical share of PatchTST IG increases in both noisy settings, with median changes of 0.152 under high magnitude noise and 0.106 under high autocorrelation (both $p < 0.001$).

Taken together, the evidence from the representative case in Fig.~\ref{fig:fig8} and the broader summaries across samples in Supplementary Fig.~\ref{fig:supp_interpretability_consistency} and Supplementary Table~\ref{tab:supp_interpretability_stats} suggest that feature substitution can be interpreted in two related ways. At the variable level, the model reduces its dependence on contaminated radiation forecasts and shifts toward historical and other more stable information sources. At the temporal level, the model no longer concentrates attribution as strongly in the future window, but instead makes greater complementary use of historical observations. This combined interpretation is consistent with the robustness patterns observed in Section~3 and the physically modulated amplification patterns discussed in Section~4.

\FloatBarrier

\section{Pareto Trade-off Analysis and Engineering Implications}

The preceding analyses indicate that model selection for PV power forecasting cannot rely on a single performance metric. Instead, engineering use requires a multi-objective trade-off among accuracy under clean conditions, robustness to diverse error mechanisms, and computational efficiency.

\subsection{Accuracy vs. Noise Magnitude Pareto Trade-off}

LightGBM consistently establishes the accuracy upper bound under ideal clean conditions. However, its steep performance deterioration under increasing noise magnitude pushes it away from the optimal Pareto frontier in volatile environments. This pattern is especially evident in Z2, Z3, and Z5. For instance, in the highly volatile Z3 region, LightGBM exhibits an Area Under the Error Curve (AUEC) with respect to noise magnitude of approximately 0.28, indicating a high cumulative error risk across diverse perturbation scenarios. In contrast, PatchTST and GRU maintain significantly lower cumulative risk profiles (AUEC $\approx$ 0.20 and 0.18, respectively), supporting their relative advantage under severe uncertainty. Deep sequence architectures, specifically PatchTST and GRU, more consistently lie near the Pareto frontier. Notably, PatchTST often forms a knee point on the trade-off curve, sacrificing only marginal accuracy under clean conditions in exchange for substantial gains in noise robustness. While N-HITS offers competitive latency, it frequently falls slightly behind the Pareto frontier in terms of absolute robustness.

\subsection{Computational Latency vs. Autocorrelation Robustness Trade-off}

High-frequency rolling forecasts in real world PV plants often require immediate local execution on edge computing devices or industrial controllers, where GPU acceleration is typically unavailable. Therefore, computational latency must be evaluated using CPU execution under a single-sample inference protocol. In this domain, LightGBM demonstrates minimal sensitivity to temporal autocorrelation, yet incurs the highest computational overhead across regions, remaining around 10--11 ms/sample in the five climate zones. This latency is still short enough for forecasting at a single site, and therefore should not be interpreted as making tree-based models impractical. Its engineering relevance instead emerges in scheduling at the grid or fleet level, where forecasts may need to be refreshed repeatedly for hundreds or thousands of distributed PV nodes and millisecond level differences per node can accumulate into a non-negligible operational burden. While MLP provides the fastest inference, it is highly vulnerable to persistent errors. PatchTST and N-HITS are highly competitive in the low latency regime while maintaining acceptable temporal robustness. GRU offers strong robustness to $\rho$, albeit at a slightly higher inference cost than PatchTST and N-HITS.

\begin{figure}[htbp!]
    \centering
    \includegraphics[width=0.98\textwidth, height=0.85\textheight, keepaspectratio]{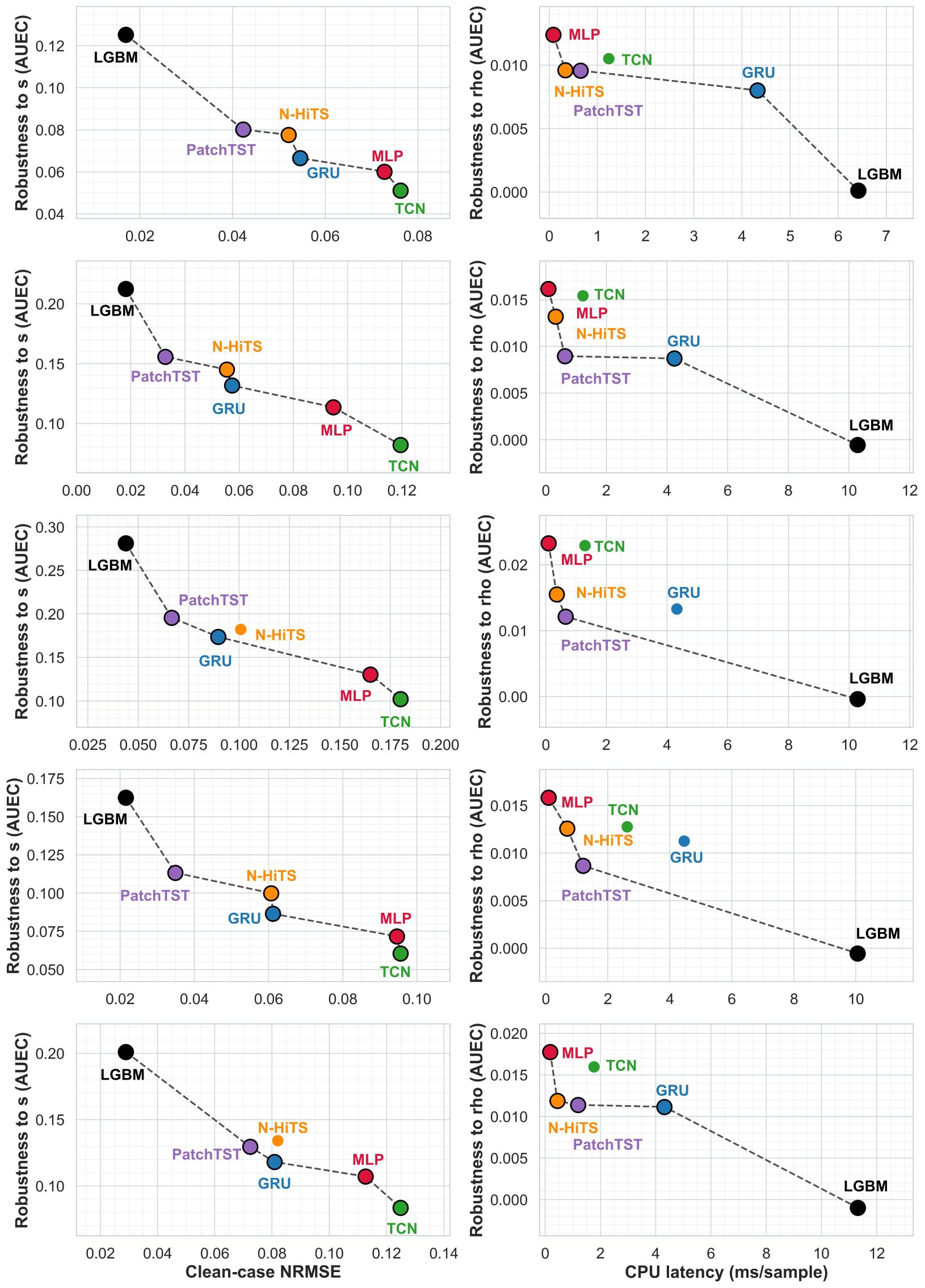}
    \caption{Comprehensive model selection and Pareto trade-off analysis. (a), (c), (e), (g), and (i) Trade-off frontiers between accuracy under clean conditions and robustness to noise magnitude across the five climate zones, where robustness is quantified by the Area Under the Error Curve (AUEC). (b), (d), (f), (h), and (j) Trade-offs between computational efficiency and robustness to error autocorrelation, also measured by AUEC. The latency metric is measured as CPU end-to-end single-sample inference time for the full 24-step forecast horizon, including feature construction when applicable.}
    \label{fig:fig9}
\end{figure}

It is important to note that the latency reported here refers to CPU end-to-end single-sample inference time for the full 24-step forecast horizon, rather than training time. The benchmark includes both feature construction and model forward time. Under this DMS protocol, LightGBM evaluates 24 regressors specific to individual forecast horizons and repeats feature engineering for each lead time, whereas the deep sequence models generate the full horizon in a single forward pass. Consequently, the latency ranking in Fig.~\ref{fig:fig9} should be interpreted as an implementation-aware deployment benchmark for this forecasting setup, rather than as a universal statement that tree-based models are intrinsically slower than deep learning models.

\subsection{Implications for Model Selection}

These multidimensional trade-offs suggest that model choice should be governed by the specific forecasting environment rather than by a universally superior architecture. If the application assumes high quality NWP inputs and prioritizes peak accuracy, LightGBM remains a highly suitable option. However, under unstable, HRRR-calibrated NWP perturbations, particularly in meteorologically active regions with frequent rolling updates, sequence models provide a better balance of performance. Overall, PatchTST appears to be the most versatile candidate in the present benchmark, offering an effective balance among baseline accuracy, temporal robustness, and computational efficiency, while GRU remains highly competitive in several difficult regions such as Z3 and Z5.

\FloatBarrier

\section{Discussion and Future Directions}

While this study systematically quantifies model robustness under dynamic NWP forecast errors, its primary contribution is methodological: it establishes a physically constrained benchmark for isolating how structured input uncertainty propagates through different forecasting architectures. The use of virtual PV power is a deliberate modeling choice rather than a surrogate claim of full field realism. By controlling the response generation process, the framework filters out confounders at the plant level, such as equipment heterogeneity, curtailment, maintenance events, and undocumented operational disturbances, allowing robustness degradation to be attributed more directly to meteorological drivers and internal model behavior.

The benchmarking scope should also be interpreted with two additional boundaries in mind. First, the comparison between LightGBM and the sequence models is based on matched information sources but not identical feature representations. It therefore reflects model family behavior under representative usage patterns rather than a strict architecture only fairness test. Second, the chronological split of 2018--2024 provides a controlled train/validation/test protocol, but it does not by itself establish full robustness across longer-term interannual climate variability.

These controlled assumptions also define the scoping boundaries of the present study. The reported results should be interpreted as controlled engineering evidence of relative model sensitivity under HRRR-calibrated forecast perturbations, rather than as a direct estimate of performance at a specific operating PV plant. This scope is intentional: by first isolating the propagation of meteorological input uncertainty, the benchmark provides a clearer basis for comparing model families before adding operational effects specific to individual sites. A natural next step is therefore to relax these boundaries in stages. First, while the framework emulates dynamic forecast errors, it currently assumes a relatively clean historical observation window. Extending the benchmark to include sensor drift, synchronization errors, and missing data anomalies would test whether the observed robustness patterns persist under imperfect measurement streams. Second, integrating field data from utility scale systems would help evaluate the same patterns when plant level effects such as inverter clipping, soiling, localized shading, and dispatch constraints are also present. Third, broader multi-year testing would help assess the stability of these comparative findings under stronger interannual climate variability and more diverse extreme weather realizations.

The perturbation design should be viewed under the same scoping logic. The present study focuses on one compact, empirically anchored parameterization so that comparisons across models, zones, and scenarios remain interpretable under a common protocol. This design choice favors methodological clarity over exhaustive coverage of the perturbation space, allowing the analysis to identify how representative HRRR-calibrated error structures propagate through different forecasting architectures. Consequently, the robustness rankings reported here should be read as stable within the tested settings, rather than as proof of invariance to all alternative choices of autoregressive structure, heteroscedastic scaling, or temperature noise specification. To probe this boundary without losing the compactness of the main experiment, we include a local sensitivity check around the representative settings in Supplementary Fig.~\ref{fig:supp_local_sensitivity} and Supplementary Table~\ref{tab:supp_local_sensitivity}. Within this tested neighborhood, the model rankings by zone remain highly similar overall, with the mean Spearman rank correlation to the base setting exceeding 0.86 in every case, although the $\rho$-local sweep in Z1 Arid and Z5 High Plains shows moderate rotation in the identity of the top-1 or top-2 model set, indicating that fine-grained model ordering in these two zones is somewhat more sensitive to the exact autocorrelation level than in the remaining zones. A full study of the perturbation space remains an important direction for future work.

Finally, expanding the perturbation framework beyond key radiation and temperature drivers to encompass fully joint multivariate NWP error structures will provide a more comprehensive stress test. Combining meteorological physics constraints or physics-informed learning with causal AI methodologies may further help future investigations move beyond case-based feature attribution toward stricter causal pathways, thereby bridging the gap between atmospheric dynamics and the engineering use of deep learning forecasting systems.

\section{Conclusion}

This study systematically investigated the robustness of PV power forecasting models against HRRR-calibrated NWP forecast errors using a physically constrained evaluation framework across five diverse U.S. climate zones. By dynamically corrupting future irradiance and temperature inputs while preserving radiation consistency through Erbs reconstruction, we characterized the vulnerabilities and adaptive response patterns of six representative models under controlled but practically motivated perturbation settings.

Three principal conclusions emerge from this analysis. First, model robustness loss is strongly nonlinear. While gradient boosted tree models establish the accuracy upper bound under ideal conditions, their performance deteriorates rapidly as forecast uncertainty increases. Conversely, deep sequence models, particularly GRU and PatchTST, show stronger resilience in medium to high noise regimes within the present benchmark setting. Second, the spatial heterogeneity of forecasting degradation is physically governed rather than purely statistical. Climate zones characterized by frequent sky state jumps or high short-term structural persistence (e.g., the Humid Continental zone) nonlinearly amplify input perturbations, linking algorithmic vulnerability directly to local atmospheric dynamics. Third, under degraded future inputs, the attribution results suggest a redistribution of predictive reliance away from unreliable meteorological forecasts toward historical and other comparatively stable information sources.

From an engineering perspective, these findings emphasize the risk of selecting forecasting models based solely on accuracy benchmarks under clean conditions. Model selection should instead be navigated through a Pareto trade-off encompassing accuracy under clean conditions, noise robustness, and computational latency under CPU execution, tailored to the specific meteorological environment and operational constraints. Ultimately, evaluating models under complex, empirically calibrated forecast error structures provides a more informative basis for comparing renewable energy forecasting systems in an inherently uncertain atmosphere.

\section*{Acknowledgements}

This research was supported by the National Natural Science Foundation of China (Grant 42450105), the Open Research Fund of the Key Laboratory of High Impact Weather (special), China Meteorological Administration (Grant 2025-G-15), and the Science and Technology Development Foundation of the Chinese Academy of Meteorological Sciences (Grant 2024KJ007). The authors also gratefully acknowledge the National Renewable Energy Laboratory (NREL) for providing the open-access meteorological datasets used in this study, and the developers and maintainers of the \texttt{pvlib} Python community for the open-source tools that supported the physical simulations.

\section*{CRediT authorship contribution statement}

Dandan Chen: Conceptualization, Methodology, Investigation, Formal analysis, Visualization, Writing - original draft.
Yan Zhao: Conceptualization, Writing - review \& editing.
Xuepeng Chen: Formal analysis, Validation, Writing - review \& editing.

\section*{Data Availability}

The meteorological driving data used in this study are publicly available from the National Renewable Energy Laboratory (NREL) National Solar Radiation Database (NSRDB; \url{https://nsrdb.nrel.gov/}). The High-Resolution Rapid Refresh (HRRR) forecast data used for residual calibration are publicly available through the NOAA Open Data Dissemination Program / AWS public archive (\url{https://registry.opendata.aws/noaa-hrrr-pds/}). The virtual photovoltaic power data were generated using the open-source Python library pvlib (\url{https://pvlib-python.readthedocs.io/}). The processed datasets and underlying derived data supporting the findings of this study are available from the corresponding author upon reasonable request.

\section*{Declaration of Competing Interest}

The authors declare that they have no known competing financial interests or personal relationships that could have appeared to influence the work reported in this paper.

\bibliographystyle{elsarticle-num}
\bibliography{references}

\clearpage

\section*{Supplementary Material}
\setcounter{figure}{0}
\renewcommand{\thefigure}{S\arabic{figure}}
\setcounter{table}{0}
\renewcommand{\thetable}{S\arabic{table}}

\begin{figure}[htbp!]
    \centering
    \includegraphics[width=0.98\textwidth]{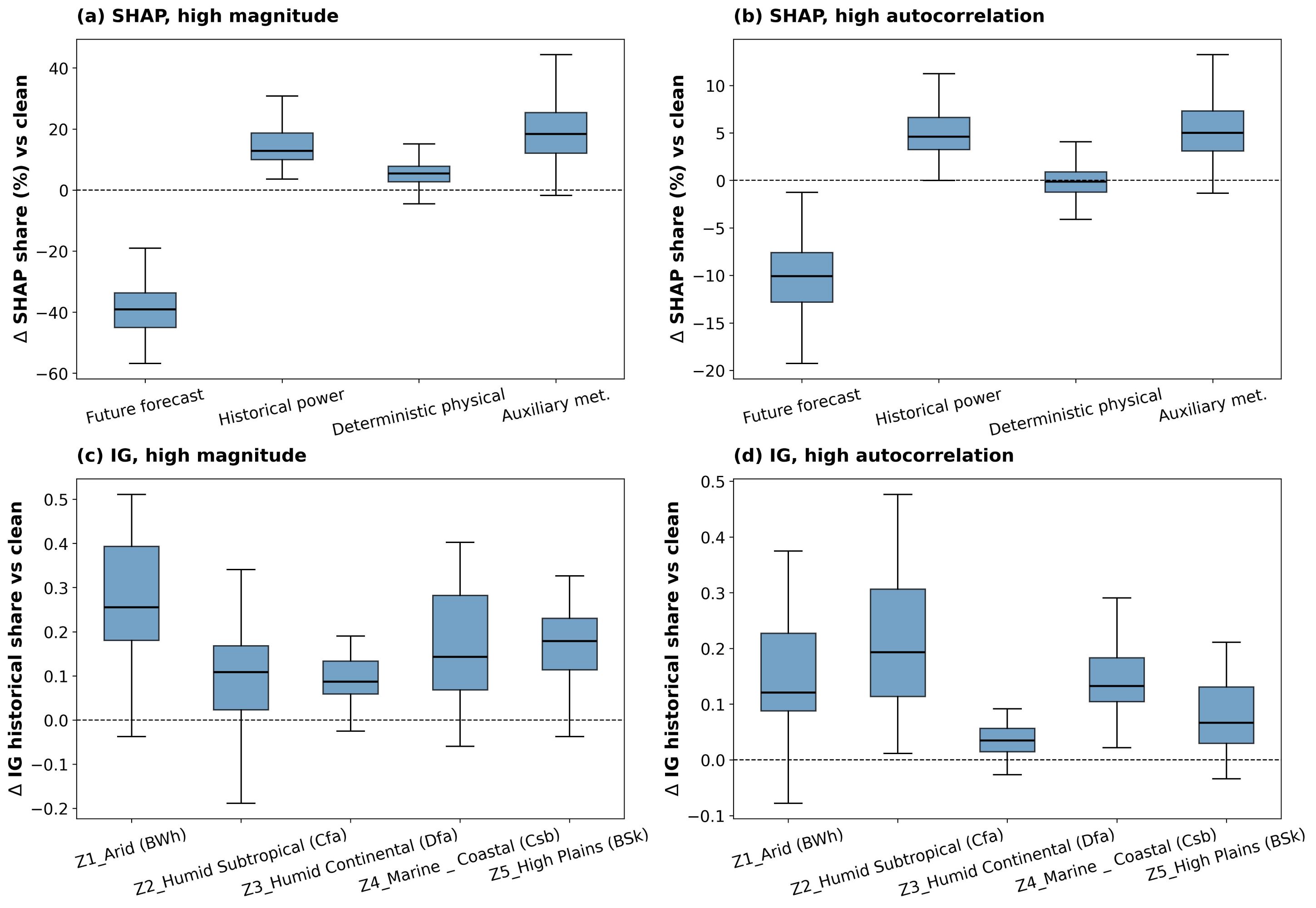}
    \caption{Consistency of the interpretability shifts at the population level across 250 paired daytime test samples from five climate zones. (a)--(b) Changes in LightGBM SHAP category shares relative to the clean baseline under the high magnitude noise and high autocorrelation scenarios, respectively. (c)--(d) Changes in PatchTST historical-window Integrated Gradients (IG) share relative to the clean baseline under the same two scenarios. The paired summaries show that future forecast feature importance decreases consistently under both noisy settings, while reliance on historical power and auxiliary meteorology tends to increase. The historical share of IG attribution also increases under both perturbation types, supporting a recurring but not fully uniform feature reallocation tendency beyond the representative case shown in Fig.~\ref{fig:fig8}.}
    \label{fig:supp_interpretability_consistency}
\end{figure}

\begin{table}[htbp]
    \centering
    \caption{Interpretability summary statistics at the population level corresponding to Supplementary Fig.~\ref{fig:supp_interpretability_consistency}. Reported values are median changes relative to the clean baseline, together with the interquartile range (IQR) and paired Wilcoxon signed-rank $p$-value across 250 paired daytime test samples.}
    \label{tab:supp_interpretability_stats}
    \scriptsize
    \setlength{\tabcolsep}{4pt}
    \resizebox{\textwidth}{!}{\begin{tabular}{llccc}
        \toprule
        Scenario & Metric & Median $\Delta$ & IQR $\Delta$ & $p$-value \\
        \midrule
        High magnitude ($\rho=0$, $s=0.3$) & Future forecast share & $-39.13$ & [$-44.94$, $-33.61$] & $<0.001$ \\
        High magnitude ($\rho=0$, $s=0.3$) & Historical power share & $12.79$ & [$9.99$, $18.76$] & $<0.001$ \\
        High magnitude ($\rho=0$, $s=0.3$) & Deterministic physical share & $5.38$ & [$2.78$, $7.81$] & $<0.001$ \\
        High magnitude ($\rho=0$, $s=0.3$) & Auxiliary meteorology share & $18.31$ & [$12.13$, $25.39$] & $<0.001$ \\
        High magnitude ($\rho=0$, $s=0.3$) & IG historical share & $0.152$ & [$0.081$, $0.234$] & $<0.001$ \\
        High autocorrelation ($\rho=0.5$, $s=0.1$) & Future forecast share & $-10.08$ & [$-12.82$, $-7.59$] & $<0.001$ \\
        High autocorrelation ($\rho=0.5$, $s=0.1$) & Historical power share & $4.61$ & [$3.26$, $6.63$] & $<0.001$ \\
        High autocorrelation ($\rho=0.5$, $s=0.1$) & Deterministic physical share & $-0.12$ & [$-1.23$, $0.90$] & $0.224$ \\
        High autocorrelation ($\rho=0.5$, $s=0.1$) & Auxiliary meteorology share & $5.01$ & [$3.11$, $7.33$] & $<0.001$ \\
        High autocorrelation ($\rho=0.5$, $s=0.1$) & IG historical share & $0.106$ & [$0.049$, $0.179$] & $<0.001$ \\
        \bottomrule
    \end{tabular}
    }
\end{table}

\begin{figure}[htbp!]
    \centering
    \includegraphics[width=0.98\textwidth]{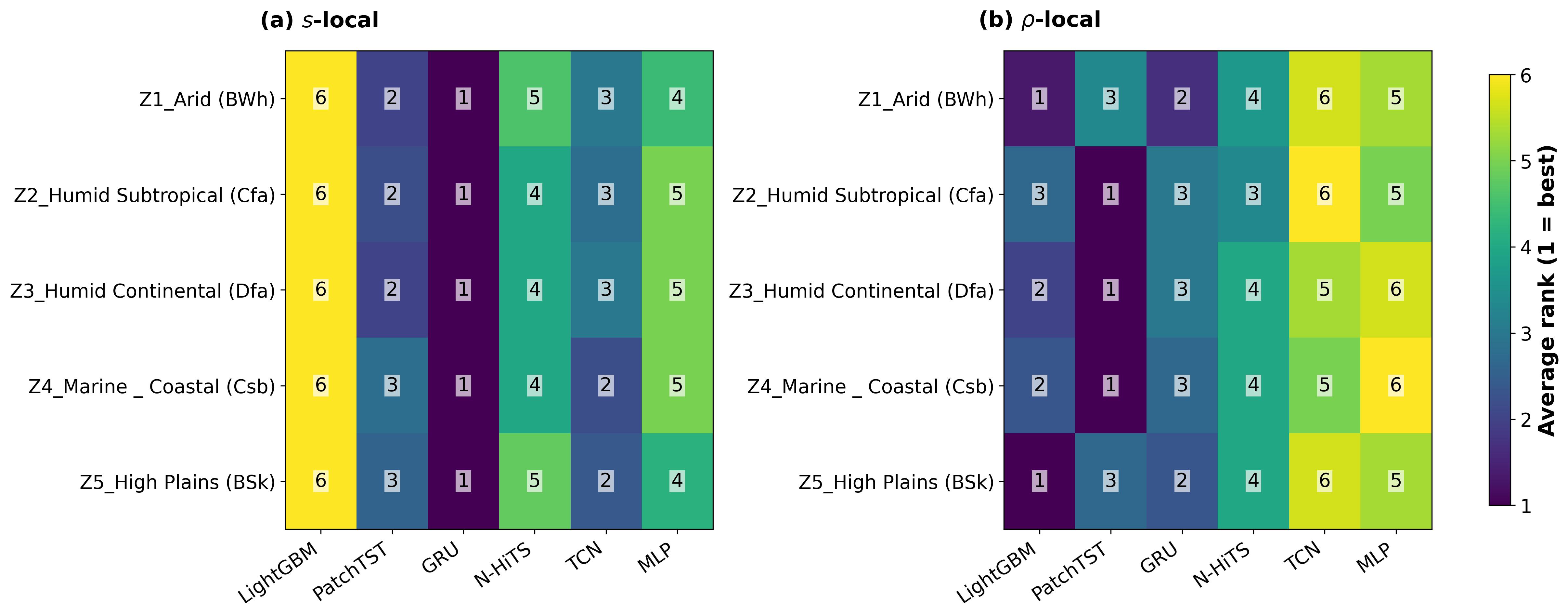}
    \caption{Compact local sensitivity check of the model rankings by zone around the two representative perturbation settings used in the manuscript. (a) Local variation around the high magnitude setting ($\rho=0$, $s=0.3$) using $s \in \{0.24, 0.28, 0.30, 0.32, 0.36\}$. (b) Local variation around the high autocorrelation setting ($\rho=0.5$, $s=0.1$) using $\rho \in \{0.3, 0.5, 0.7\}$ at fixed $s=0.1$. Cell values denote average rank across the tested neighborhood (1 = best). Aggregate rank stability statistics are reported in Supplementary Table~\ref{tab:supp_local_sensitivity}.}
    \label{fig:supp_local_sensitivity}
\end{figure}

\begin{table}[htbp]
    \centering
    \caption{Summary by zone of the compact local sensitivity check in Supplementary Fig.~\ref{fig:supp_local_sensitivity}. Top-1 and Top-2 stability fractions are computed relative to the base setting within each local neighborhood.}
    \label{tab:supp_local_sensitivity}
    \scriptsize
    \setlength{\tabcolsep}{4pt}
    \resizebox{\textwidth}{!}{\begin{tabular}{llcccccc}
        \toprule
        Sweep & Zone & Base & Top model & Neighbors & Top-1 stable & Top-2 stable & Mean rank corr. \\
        \midrule
        $s$-local & Z1 Arid & $s=0.3$ & GRU & 4 & 1.00 & 1.00 & 0.97 \\
        $s$-local & Z2 Humid Subtropical & $s=0.3$ & GRU & 4 & 1.00 & 0.75 & 0.99 \\
        $s$-local & Z3 Humid Continental & $s=0.3$ & GRU & 4 & 1.00 & 1.00 & 1.00 \\
        $s$-local & Z4 Marine/Coastal & $s=0.3$ & GRU & 4 & 1.00 & 0.75 & 0.99 \\
        $s$-local & Z5 High Plains & $s=0.3$ & GRU & 4 & 1.00 & 0.50 & 0.96 \\
        $\rho$-local & Z1 Arid & $\rho=0.5$ & LightGBM & 2 & 0.50 & 1.00 & 0.89 \\
        $\rho$-local & Z2 Humid Subtropical & $\rho=0.5$ & PatchTST & 2 & 1.00 & 0.50 & 0.86 \\
        $\rho$-local & Z3 Humid Continental & $\rho=0.5$ & PatchTST & 2 & 1.00 & 1.00 & 0.97 \\
        $\rho$-local & Z4 Marine/Coastal & $\rho=0.5$ & PatchTST & 2 & 1.00 & 0.50 & 0.97 \\
        $\rho$-local & Z5 High Plains & $\rho=0.5$ & LightGBM & 2 & 1.00 & 0.00 & 0.91 \\
        \bottomrule
    \end{tabular}
    }
\end{table}

\end{document}